\documentclass[a4paper,10pt,twoside]{cpc-hepnp}
\usepackage{amssymb}
\usepackage{}

\usepackage{multicol}
\usepackage{graphicx}
\usepackage{booktabs}
\usepackage{amssymb,bm,mathrsfs,bbm,amscd}
\usepackage[tbtags]{amsmath}
\usepackage{lastpage}

\begin{document}

\fancyhead[c]{\small Submitted to 'Chinese Physics C'
} \fancyfoot[C]{\small 010201-\thepage}

\footnotetext[0]{Received 12 November 2013}

\title{Feasibility Study on SRF Cavity Improvement by Doping Method\thanks{supported by National
Natural Science Foundation of China (Grant No 11175008)}}

\author{JIANG Tao
\quad HE Fei-Si
\quad JIAO Fei$^{1)}$\email{fjiao@pku.edu.cn}\\
\quad HE Fa
\quad LU Xiang-Yang
\quad ZHAO Kui
}
\maketitle

\address{State Key Laboratory of Nuclear Physics and Technology, Peking University, Beijing 100871, China\\}

\begin{abstract}
  In this article a new idea is proposed by PKU group to improve the RF superconductive property of niobium cavity. Rare earth elements like scandium and yttrium are doped into ingot niobium during the smelting processes. A series of experiments have been done since 2010, and the preliminary testing results show that the magnetic property of niobium materials has been changed with different doping elements and proportions while the superconductive transition temperature almost does not change. Then a large ingot niobium doped with scandium was melted. RRR and mechanical properties were measured. This method may increase the superheating magnetic field and decrease surface resistance of niobium so as to improve the performance of niobium cavity which is a key component of SRF accelerators. A single-cell cavity made of Sc-doped niobium has been fabricated.
\end{abstract}



\footnotetext[0]{\hspace*{-3mm}\raisebox{0.3ex}{$\scriptstyle\copyright$}2013
Chinese Physical Society and the Institute of High Energy Physics
of the Chinese Academy of Sciences and the Institute
of Modern Physics of the Chinese Academy of Sciences and IOP Publishing Ltd}%

\begin{multicols}{2}

\section{Introduction}
Superconductive RF (SRF) accelerator is widely used in X-ray FEL, ADS Spallation Neutron Source and the light source of ERL. The future ILC also prefers to use the SRF technology. The niobium cavity is the key part of the SRF accelerator. After about 40 years of development, great progress has been made on the cavity. Higher performance of cavities, especially in the area of accelerating gradient ($E_{\mathrm{acc}}$) and quality factor ($Q_{0}$), is required for beam acceleration of high energy, current and quality, which will promote the development of large scientific facilities. Research has begun recently in the area of improving the property of cavity material and structure.

The accelerating gradient has been pushed to the theoretical limit. For a certain type of cavity, $E_{\mathrm{acc}}$ is suppressed by peak surface magnetic field which is related to the superheating critical magnetic field ($H_{\mathrm{sh}}$). So increasing the critical magnetic field of niobium (both $H_{\mathrm{c1}}$ and $H_{\mathrm{c2}}$) will raise the $H_{\mathrm{sh}}$ in order to increase $E_{\mathrm{acc}}$. The rf surface resistance ($R_{\mathrm{s}}$), which is one of the limitation of the performance of the cavity, consists of two parts, BCS resistance ($R_{\mathrm{BCS}}$) and residual resistance ($R_{0}$). The former part is related to the normal electrons in superconductor. The latter part can be controlled by the manufacture process in the order of $n\Omega$. There are various means to improve $E_{\mathrm{acc}}$ and $R_{\mathrm{s}}$, such as large grain niobium\cite{L}, multifilm\cite{G} and new superconductive material.

A new idea\cite{1} is proposed by the group of Peking University (PKU) in this article that the property of the type-II superconductor niobium, such as critical magnetic field and surface resistance, can be improved by introducing impurity into ingot niobium during the smelting process. Since the end of 2010, a lot of theoretical and experimental research has been done by people of PKU and Ningxia Orient Tantalum Industry Co., Ltd. (OTIC). Firstly, a series of smelting processes were conducted to produce small-size niobium samples with different proportions of scandium, yttrium and lanthanum. The component, transition temperature and magnetization curve of the new materials were measured. The results agreed well with our expectation.Then OTIC melted the first large-size ingot doped with scandium. Impurity content, residual resistivity ratio (RRR) and mechanical property of this ingot were measured, which satisfied the demand of SRF cavity. On this basis, the first cavity made of this new material from the same ingot has been fabricated. The welding effect on the impurity content was studied. Vertical test was taken just after simple polishing treatment. The result reveals preliminarily that there is no bad defect in this niobium cavity doped with scandium. More post-treatment (such as high-temperature baking and electropolishing) and further test will be carried out.

\section{Effect of Impurity}
\subsection{Effect on $H_{\mathrm{c1}}$ by Doped Impurity}
In order to understand the dependence of $H_{\mathrm{c1}}$ on the electron mean free path $l$, we tried to find out the relationship between $H_{\mathrm{c1}}$ and $l$ by using the result of Ginzburg-Landau theory in £¢dirty£¢ limit.

From Ginzburg-Landau theory, when Ginzburg-Landau parameter $\kappa \gg  1$ we can have\cite{6.3}
\begin{eqnarray}
 \label{eq1}
  H_{\mathrm{c1}}(T)=\frac{1}{\sqrt{2}}H_{\mathrm{c}}(T)\frac{1}{\kappa}\ln\kappa,
\end{eqnarray}
\begin{eqnarray}
 \label{eq2}
  H_{\mathrm{c}}(T)=\frac{\phi_{\mathrm{0}}}{\sqrt{2}\pi\mu_{\mathrm{0}}\xi(T)\lambda(T)},
\end{eqnarray}
 where $\xi(T)$ is the coherence length, $\lambda(T)$ is the penetration depth, $\phi_{0}$ is the magnetic flux quantum, $\mu_{0}$ is the permeability of vacuum. At the £¢dirty£¢ limit ($l<\xi_{\mathrm{0}}$), Gorkov gave\cite{6.5}
\begin{eqnarray}
 \label{eq3}
  \xi(T)=0.855(\xi_{\mathrm{0}}l)^{1/2}\left[\frac{T_{\mathrm{c}}}{T_{\mathrm{c}}-T}\right]^{1/2}
\end{eqnarray}
\begin{eqnarray}
 \label{eq4}
  \lambda(T)=0.64\lambda_{\mathrm{L}}(0)\left[\frac{\xi_{\mathrm{0}}}{l}\frac{T_{\mathrm{c}}}{T_{\mathrm{c}}-T}\right]^{1/2}
\end{eqnarray}
\begin{eqnarray}
 \label{eq5}
  \kappa=0.75\frac{\lambda_{\mathrm{L}}(0)}{l}
\end{eqnarray}
where $\lambda_{\mathrm{L}}(0)$ is the penetration depth at $0 K$, $\xi_{0}$ is the coherence length of pure superconductor. At $0 K$ in the £¢dirty£¢ limit ($l<\xi_{\mathrm{0}}$), there is\cite{6.8}
\begin{eqnarray}
 \label{eq6}
  \lambda_{\mathrm{L}}(0)=\lambda_{\mathrm{L}}\sqrt{1+\frac{\xi_{\mathrm{0}}}{l}}
\end{eqnarray}
where $\lambda_{\mathrm{L}}$ is the penetration depth of pure superconductor at $0 K$, a constant related to the material property. Substituting  Eq.~(\ref{eq2})$\sim$ Eq.~(\ref{eq6}) into Eq.~(\ref{eq1}), we can get
\begin{eqnarray}
  \frac{H_{\mathrm{c1}}}{\frac{\phi_{\mathrm{0}}}{0.82\pi\mu_{\mathrm{0}}\lambda_{\mathrm{L}}^{2}}\frac{T_{\mathrm{c}}-T}{T_{\mathrm{c}}}} &=& \frac{l/\xi_{\mathrm{0}}}{1+\xi_{\mathrm{0}}/l}\ln[\frac{0.75}{l}\lambda_{\mathrm{L}}\sqrt{1+\xi_{\mathrm{0}}/l}]\nonumber\\
   &=& \frac{l/\xi_{\mathrm{0}}}{1+\xi_{\mathrm{0}}/l}\ln[0.75\frac{\lambda_{\mathrm{L}}}{\xi_{\mathrm{0}}}\frac{\xi_{\mathrm{0}}}{l}\sqrt{1+\xi_{\mathrm{0}}/l}]\nonumber\\
   &=& \frac{l/\xi_{\mathrm{0}}}{1+\xi_{\mathrm{0}}/l}\ln[0.75\kappa_{\mathrm{0}}\frac{\xi_{\mathrm{0}}}{l}\sqrt{1+\xi_{\mathrm{0}}/l}],
\end{eqnarray}
where $\kappa_{\mathrm{0}}=\lambda_{\mathrm{L}}/\xi_{\mathrm{0}}$ is the Ginzburg-Landau parameter of pure superconductor.

The value of $\kappa_{\mathrm{0}}$ does not effect the tendency of $H_{\mathrm{c1}}$ over $l$. Here $\kappa_{\mathrm{0}}=5$ is adopted. Fig.~\ref{fig0} shows the dependence of $H_{\mathrm{c1}}$ on the ratio of $l$ and constant $\xi_{\mathrm{0}}$.  This manifests that certain proportion of impurity will raise $H_{\mathrm{c1}}$ of type-II superconductor.

\begin{center}
  \includegraphics[width=8.5cm]{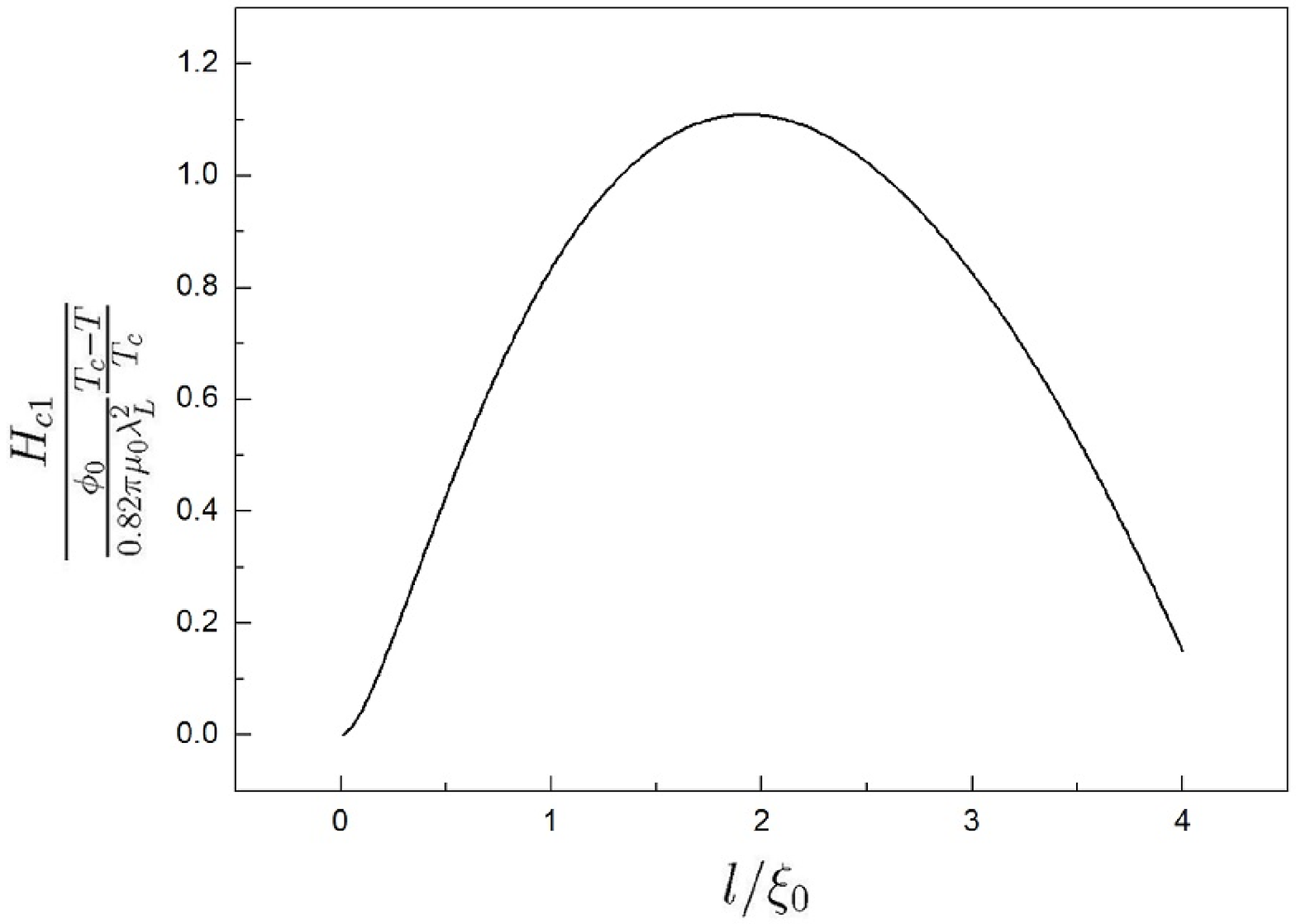}
  \figcaption{\label{fig0} The dependence of $H_{\mathrm{c1}}$ on the electron mean free path.When superconductor is pure, $l$ is much longer than $\xi_{\mathrm{0}}$, corresponding to the right end of the curve. As the impurity in superconductor increases, $l$ is shorten, and $H_{\mathrm{c1}}$ will raise up to the maximum, then drop down again.}
\end{center}

\subsection{Effect on Rs by Doped Impurity}
The RF BCS surface resistance for a certain angular frequency can be write as\cite{2}
\begin{eqnarray}
 \label{eq8}
 R_{\mathrm{BCS}}=A\lambda_{\mathrm{L}}^{3}(0)\sigma_{\mathrm{n}}\frac{\omega^{2}}{T}e^{-\frac{\Delta}{k_{\mathrm{B}}T}},
\end{eqnarray}
where $A$ is a constant which depends weakly on the material, $\omega$ is the RF angular frequency, $\Delta$ is energy gap, $k_{\mathrm{B}}$ is Boltzmann constant, $\sigma_{\mathrm{n}}$ is normal electrical conductivity, given by
\begin{eqnarray}
 \label{eq9}
 \sigma_{\mathrm{n}}=\frac{n_{\mathrm{c}}e^{2}l}{mv_{\mathrm{F}}},
\end{eqnarray}
where $n_{\mathrm{c}}$ is density of normal electrons, $m$ is effective electron mass, $v_{\mathrm{F}}$ is the Fermi velocity. From Eq.~(\ref{eq8}), Eq.~(\ref{eq9}) and Eq.~(\ref{eq6}), we can get
\begin{eqnarray}
 R_{\mathrm{BCS}}=A\lambda_{\mathrm{L}}^{3}\xi_{0}\frac{n_{\mathrm{c}}e^{2}}{mv_{\mathrm{F}}}\frac{\omega^{2}}{T}e^{-\frac{\Delta}{k_{\mathrm{B}}T}}(1+\frac{\xi_{0}}{l})^{3/2}\frac{l}{\xi_{0}}.
\end{eqnarray}
The electron mean free path dependence of $R_{\mathrm{BCS}}$ is given as following:
\begin{eqnarray}
 \frac{R_{\mathrm{BCS}}}{A\lambda_{\mathrm{L}}^{3}\xi_{0}\frac{n_{\mathrm{c}}e^{2}}{mv_{\mathrm{F}}}\frac{\omega^{2}}{T}e^{-\frac{\Delta}{k_{\mathrm{B}}T}}}=(1+\frac{\xi_{0}}{l})^{3/2}\frac{l}{\xi_{0}}.
\end{eqnarray}

Fig.~\ref{fig01} is the dependence of $R_{\mathrm{BCS}}$ on $l$. It shows that $R_{\mathrm{BCS}}$ will reach a minimum with a certain component of impurity.

\begin{center}
  \includegraphics[width=8.5cm]{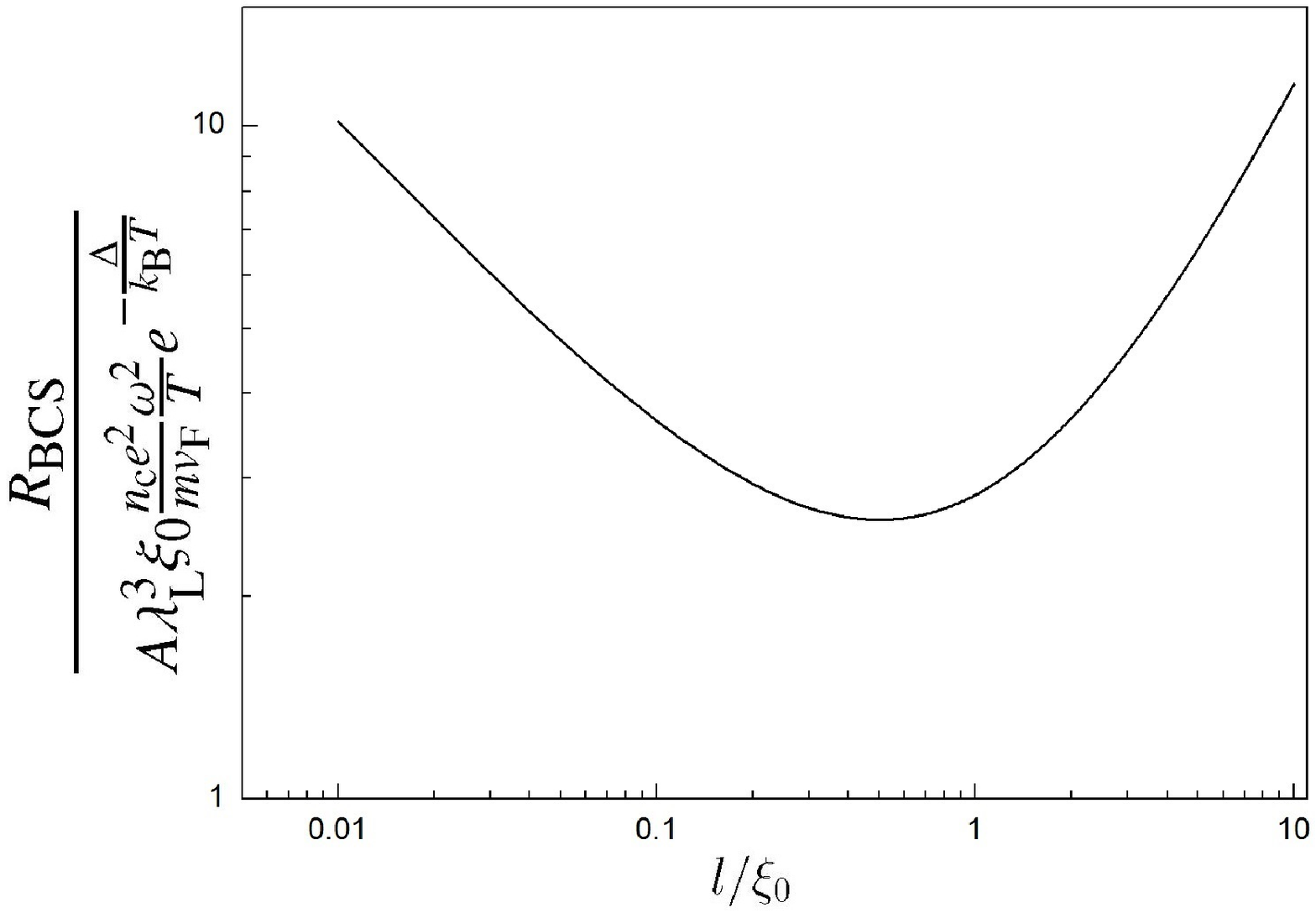}
  \figcaption{\label{fig01} The dependence of $R_{\mathrm{BCS}}$ on the electron mean free path.When superconductor is pure, $l$ is much longer than $\xi_{\mathrm{0}}$, corresponding to the right end of the curve. As the impurity in superconductor increases, $l$ is shorten, and $R_{\mathrm{BCS}}$ will go down to the minimum, then raise up again.}
\end{center}

\section{Samples Experiment}
\subsection{Samples Preparation}
Rare earth elements scandium and yttrium, which were non-magnetic, were chosen as impurities. As a comparison, we also study niobium doped with lanthanum, a magnetic element. A series of smelting processes were performed to produce small-size niobium samples with different impurity elements and proportions. Pure niobium was also prepared by the same processes as a comparison.

\subsection{Sample Test}
The compositions of small samples were analyzed by mass spectrometer (Agilent ICP-MS 7500cs). The magnetization of the samples at different temperatures and external fields were tested by the Magnetic Property Measurement System (MPMS-XL, Quantum Design).

\subsection{Results \& Discussion}
The composition of impurity in doped niobium was analyzed by a mass spectrometer. The results agreed with the quantity of elements put into the furnace, which showed that the doping composition could be controlled in the melting process.

\subsubsection{Effect on $T_{\mathrm{c}}$}
R-T and M-T curves of pure niobium and niobium doped with different elements were measured. The transition temperatures of samples doped with impurity are nearly the same as the temperature of pure niobium sample separately in the R-T and M-T testing. These curves show that the impurity does not change the superconductive transition temperature\cite{1}.

\subsubsection{Effect on $H_{\mathrm{c1}}$}
Fig.~\ref{fig3} shows the M-H curves of pure niobium and niobium doped with different elements. The magnetic field at which the M-H curve becomes nonlinear is the $H_{\mathrm{c1}}$ of the material. The linear part of M-H curve is fitted with a one-order polynomial, and the part from beginning to the minimum is fitted with a five-order polynomial. The value of $H_{\mathrm{c1}}$ can be obtained by comparing the two fitting curves. Compared to pure niobium, the Hc1 of niobium doped with non-magnetic impurities (scandium and yttrium) raise by $100 Oe$ and $200 Oe$ respectively, while that of niobium doped with magnetic impurity (lanthanum) is reduced by nearly $100 Oe$.
\begin{center}
  \includegraphics[width=8cm]{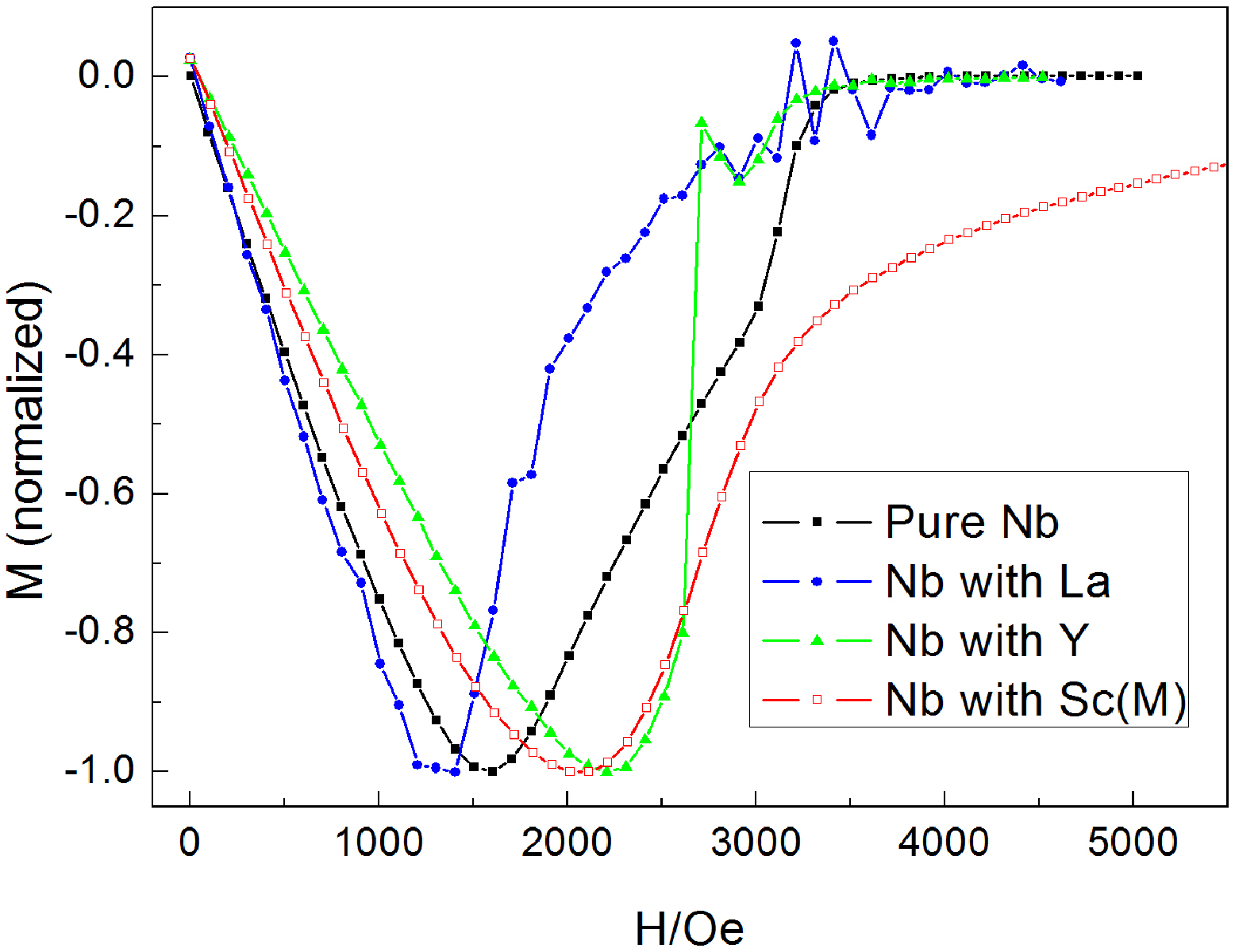}
  \figcaption{\label{fig3} (color online) M-H curve of pure Nb and Nb doped by different elements at $T=4.0 K$}
\end{center}

\subsubsection{Effect on $H_{\mathrm{c2}}$}
Fig.~\ref{fig5} shows R-T curves of another niobium sample doped with low proportion of scandium in different external magnetic fields. It is revealed that $H_{\mathrm{c2}}$ of niobium with low proportion of scandium is up to $1.1 \times 10^{4} Oe$.
\begin{center}
  \includegraphics[width=8cm]{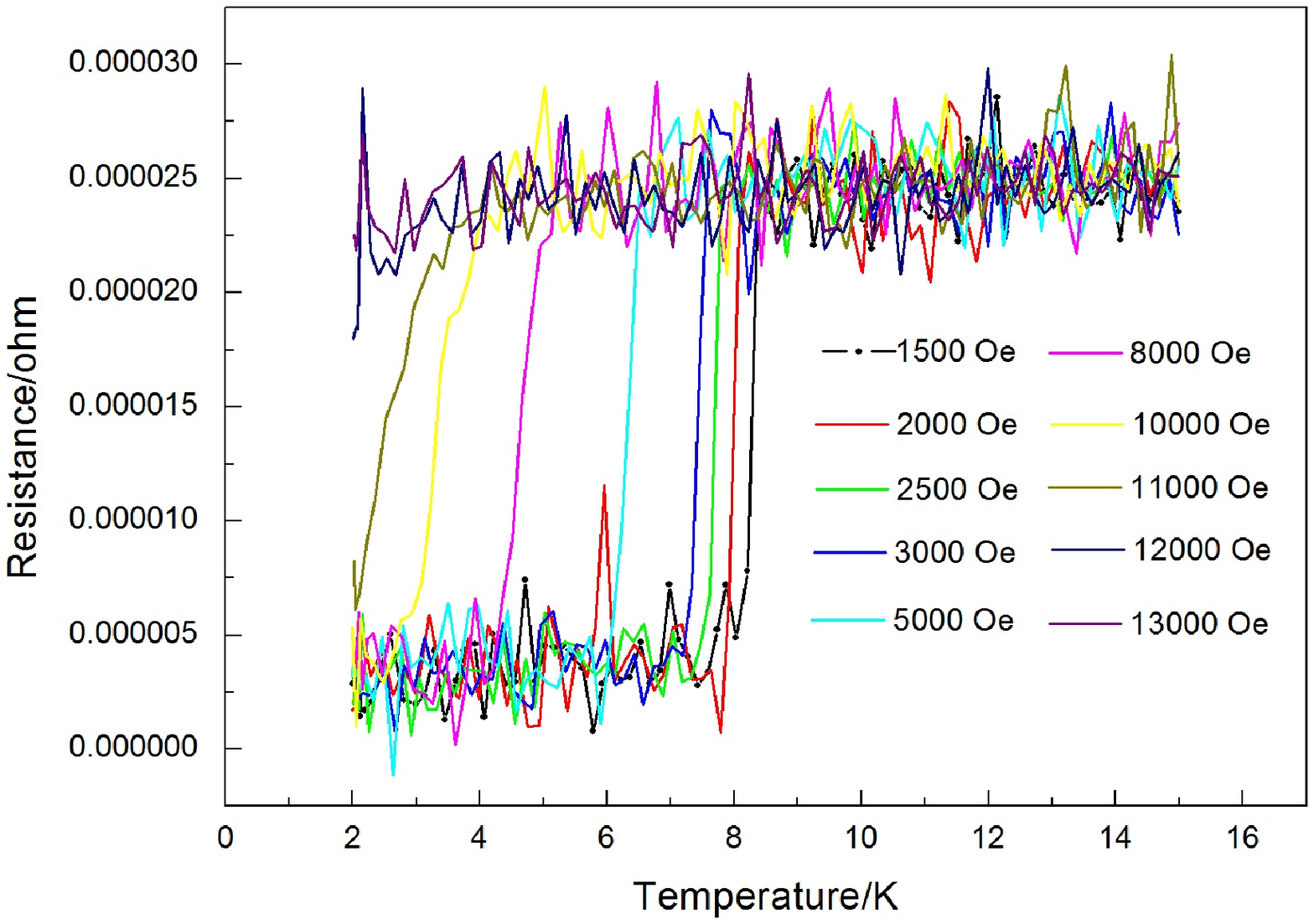}
  \figcaption{\label{fig5} (color online) R-T curves of Nb doped with low proportion of Sc in different external magnetic fields.}
\end{center}

\section{Cavity Testing}
\subsection{Preparation \& Testing}
After the tests of small samples, a large-size niobium ingot (Fig.~\ref{fig006}) doped with scandium was melted in OTIC so that samples with enough size can be prepared for RRR and mechanical properties measurement. Magnetization curves of two samples selected at random from the niobium ingot were measured.
\begin{center}
  \includegraphics[width=8cm]{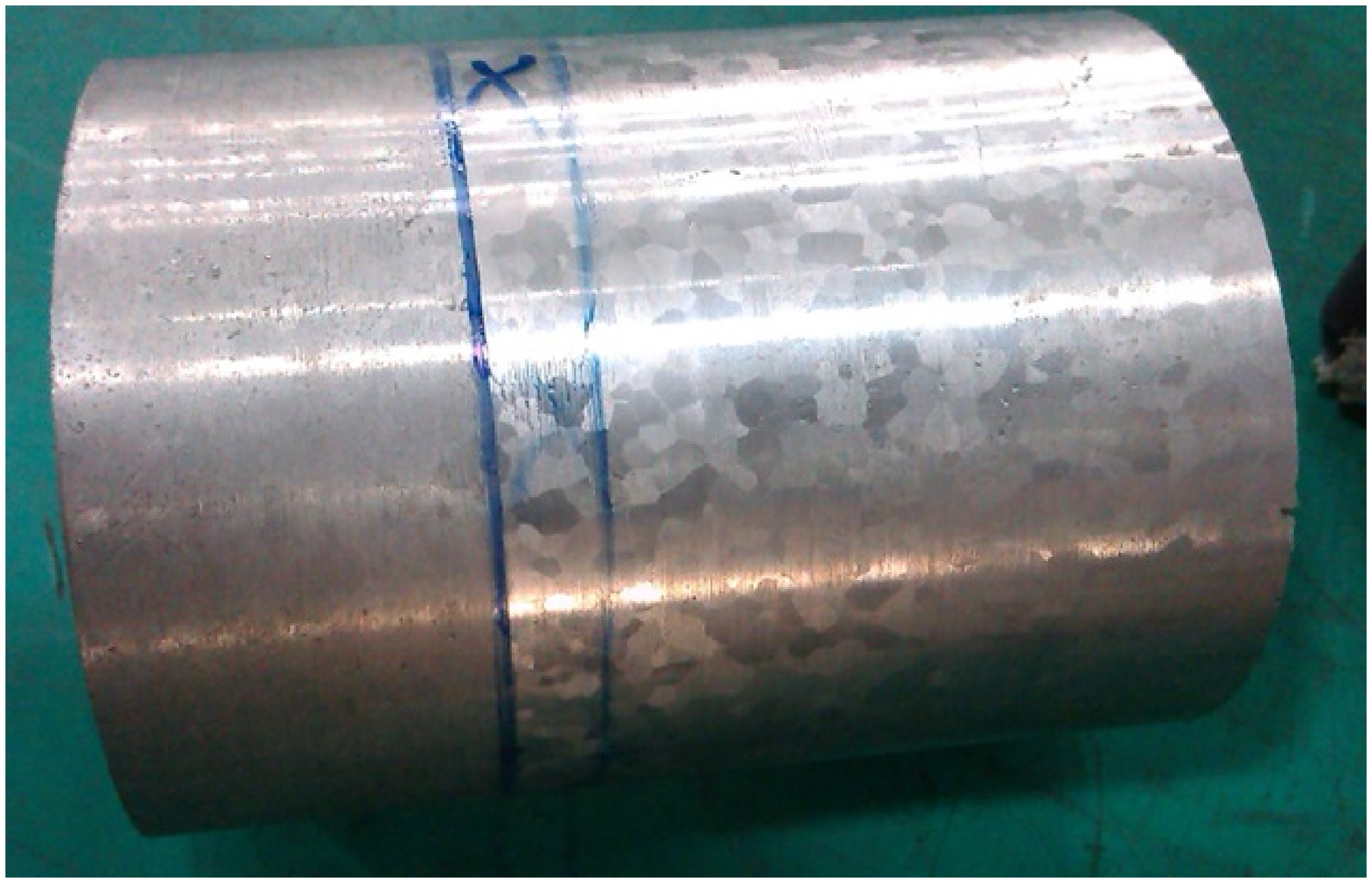}
  \figcaption{\label{fig006} (color online) Nb ingot doped with Sc melted by OTIC, nearly 20 kg. Left part is Nb-Ti alloy base, middle part is with too many defects, and right part is the Nb-Sc alloy.}
\end{center}

A $1.3 GHz$ Tesla-type single-cell cavity (Fig.~\ref{fig9.5}) was fabricated using the doped niobium sheets from the same ingot. Vertical test was carried out just after simple polishing.

\begin{center}
  \includegraphics[width=8cm]{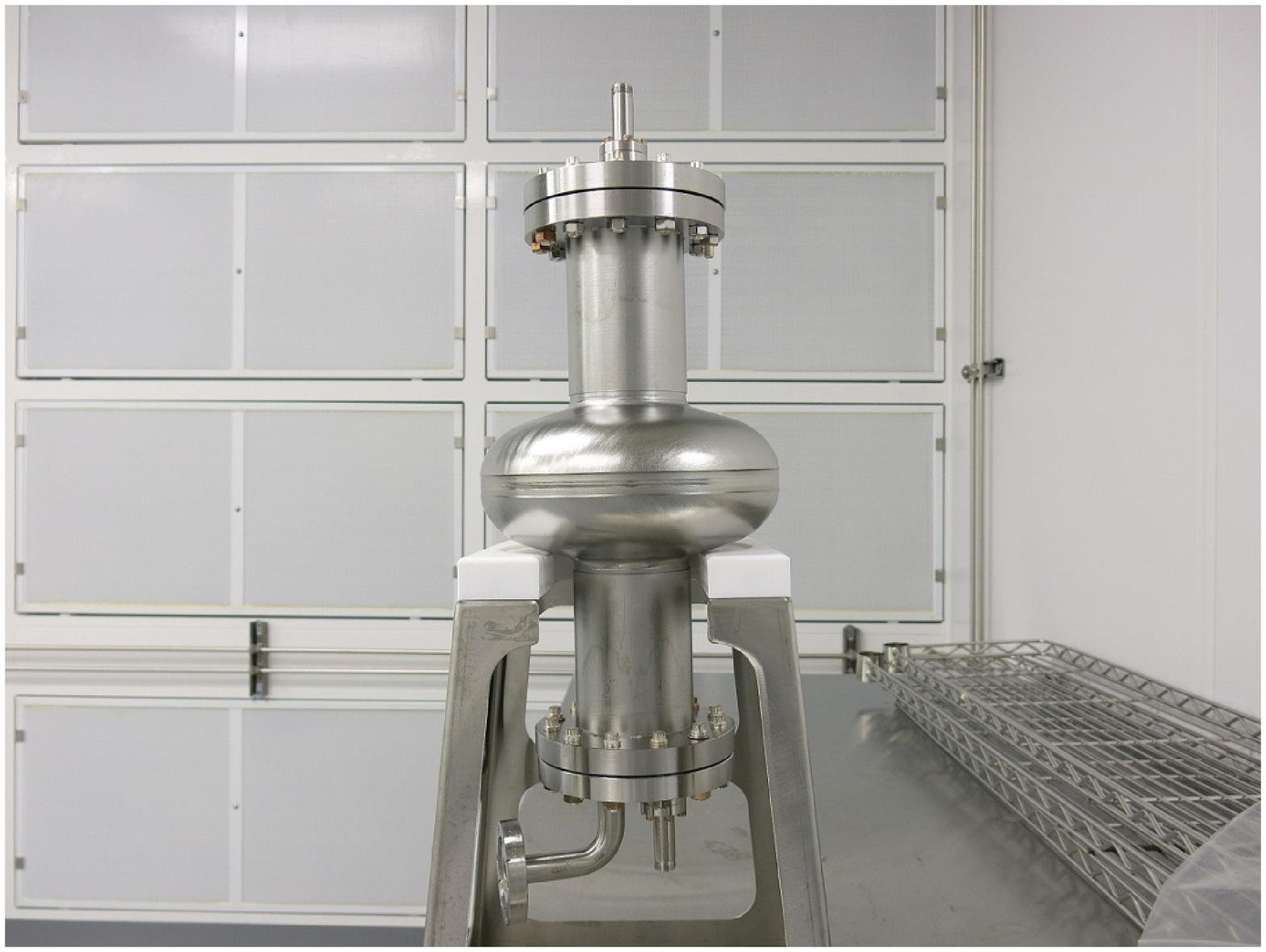}
  \figcaption{\label{fig9.5} (color online) The $1.3 GHz$ Tesla-type single-cell cavity doped with Sc}
\end{center}

\subsection{Results \& Discussion}
The results of RRR and mechanical properties of the large-size niobium ingot doped with scandium are presented in Table~\ref{tab1}. The mechanical properties satisfied the demand for SRF cavity (from CEBAF).

\begin{center}
  \tabcaption{ \label{tab1}  RRR and mechanical properties of Nb doped with Sc}
  \footnotesize
  \begin{tabular*}{80mm}{c@{\extracolsep{\fill}}cccc}
    \toprule RRR & Yield strength &  Breaking strength  & Elongation \\
    \hline
    127 & $210 MPa$ & $312.2 MPa$ & $38.7\%$ \\
    \bottomrule
  \end{tabular*}
\end{center}

M-H curves of two samples selected at random from the niobium ingot were measured. The respective $H_{\mathrm{c1}}$ from the M-H curves are $1812 Oe$ and $1811 Oe$, $500 Oe$ higher than that of pure niobium.

The Tesla-type cavity was vertical tested at $2 K$. The $Q_{0}$ was $1.3\times 10^{10}$ when the accelerating gradient was $13 MV/m$ (Fig.~\ref{fig9}). The geometry factor of this cavity is $270 \Omega$ calculated by superfish. The surface resistance $R_{\mathrm{s}}$ can be estimated to be $20.8 n\Omega$. The results reveal that at least there is no obstacle in function or craft for the application of impurity-doping in SRF niobium cavity.

\begin{center}
  \includegraphics[width=9cm]{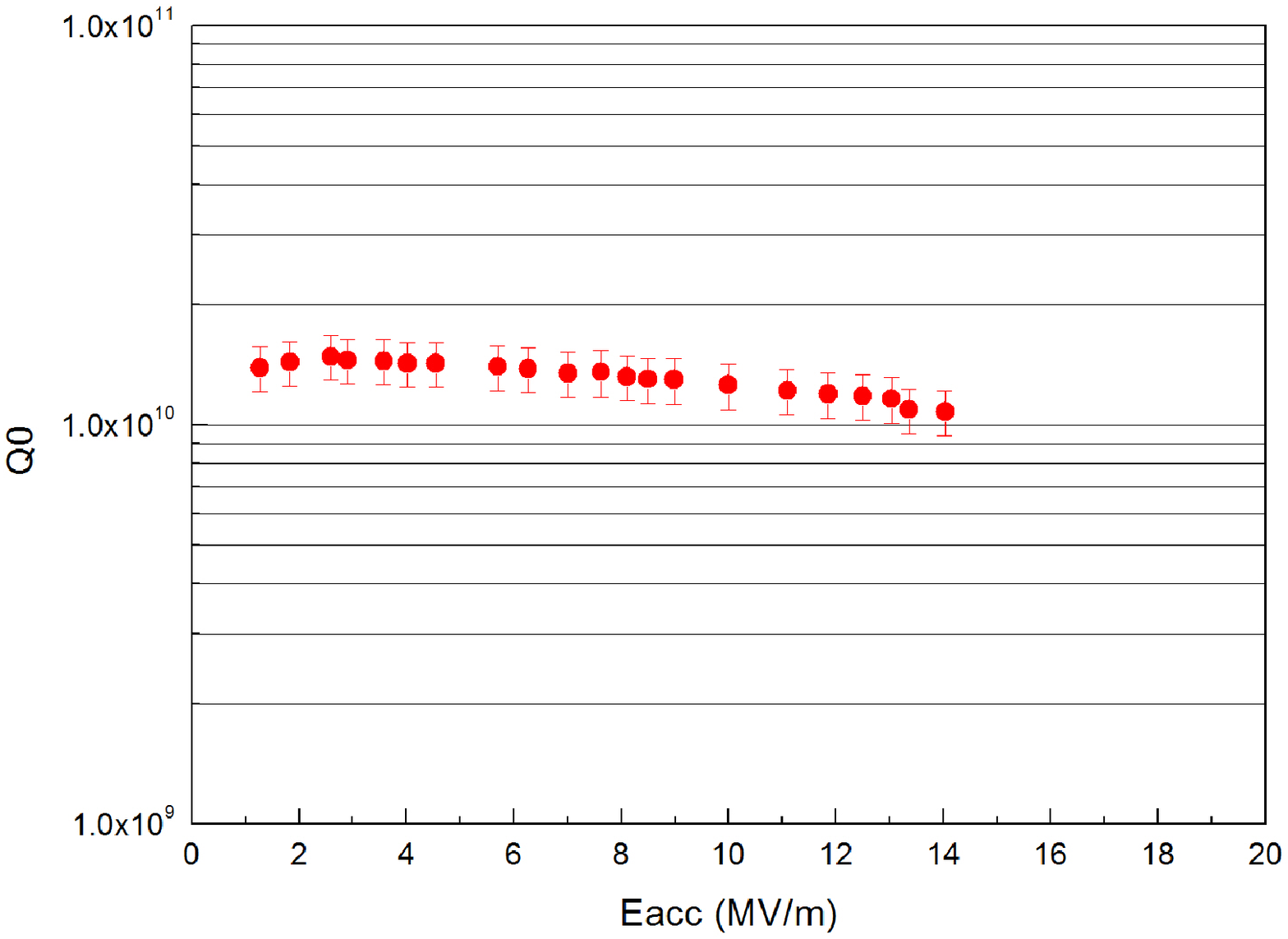}
  \figcaption{\label{fig9} (color online) The performance of the $1.3 GHz$ Tesla-type single-cell cavity doped with Sc}
\end{center}

The effect of electron-beam weld (EBW) on the impurity component was studied. A niobium sample from the ingot was welded by EBW. The M-H curves of two samples in the weld line were measured. The $H_{\mathrm{c1}}$ were reduced to $1318 Oe$ and $1317 Oe$. It indicated that the EBW purified the doped material at the equator, decreasing the component of scandium at the welded part. This will reduce the critical magnetic field of doped niobium, then lower the peak surface magnetic field of the cavity.

\section{Conclusion}
A new method is proposed in this article that doping with non-magnetic rare earth elements can change superconductive properties of niobium, so it may be a potential way to improve the performance of SRF cavity. A series of experiments have been carried out. The results show that the magnetic property of Type-II superconductor niobium is modified by the impurity of rare earth elements, improved by the non-magnetic element and degenerated by magnetic element. The transition temperature is not affected by doping. We melted the first large-size ingot of impurity-doped niobium. The mechanical performance of scandium-doped niobium satisfied the demand for SRF cavity. The first doped niobium cavity of $1.3 GHz$ Tesla type single cell was made and tested preliminarily. It reveals the feasibility in property and craft to fabricate a cavity using doped niobium.

As the effect of EBW shows, the impurity composition in the welding part is reduced. The scandium complement is in process. After regular cavity treatment, the single-cell cavity will be vertically tested.
\\

\acknowledgments{We would like to express our gratitude to H.Y. Zhao, Y.S. You and L. Chen in OTIC for their skilled melting technology and professionalism, R. Rimmer and P. Kneisel in Jlab for the cavity testing, Prof. Z.Z. Gan in PKU for evaluating the feasibility of our idea, Prof. G.F. Sun at University of Science and Technology Beijing for advice in melting, L. Lin, Dr. J.K. Hao and S.W. Quan in PKU for their support in preparation of the samples and cavity.}

\end{multicols}

%\vspace{10mm}

\vspace{-1mm}
\centerline{\rule{80mm}{0.1pt}}
\vspace{2mm}

\begin{multicols}{2}

\end{multicols}

\clearpage

\end{document}